\documentstyle[aps,prl,epsfig,multicol]{revtex}
\input epsf

\begin{document}
\draft


\title{Electronic Structure of the $YH_{3}$ Phase from Angle-Resolved 
Photoemission Spectroscopy}

 \author{J. Hayoz\cite{pra}, C. Koitzsch, M. Bovet, D. Naumovi\'{c}, L. 
 Schlapbach,  P. Aebi}
\address{D\'epartement de Physique, Universit\'e de Fribourg, P\'erolles, 
CH-1700 Fribourg, Switzerland}

\maketitle
\begin{abstract}      
Yttrium can be loaded with hydrogen up to high concentrations causing 
dramatic structural and electronic changes of the host lattice.  We 
report on angle-resolved photoemission experiments of the $Y$ trihydride 
phase.  
Most importantly, we find the absence of metal $d$-bands at 
the Fermi level and a set of flat, $H$-induced bands located at much 
higher binding energy than predicted, indicating an increased 
electron affinity at $H$ sites.
\end{abstract} 
\pacs{73.61.-r, 71.20.-b, 79.60.-i, 71.30.+h, 71.20Eh} 

\begin{multicols}{2}
\narrowtext	
%
%
Recently, switchable optical properties of metal hydrides at ambient 
pressures and temperatures have attracted strong interest.\cite{huiberts}  
For trivalent $Y$, for instance, up to three hydrogen 
atoms can be absorbed.  The dihydride is even a better metal than $Y$ 
itself but during the transition to the trihydride phase it turns from 
shiny metallic to transparent and insulating.  
With hydrogen a proton and an electron are introduced to 
the metal host.  This results in doping the host material.
Different models have been proposed to explain this spectacular 
metal-insulator transition.  However, the behavior of $H$ in such 
hydrides is still under debate.

State-of-the-art \emph{ab-initio} local density approximation 
(LDA) calculations do not reproduce the optical gap necessary to 
explain the transparent state in the trihydride phase unless 
additional symmetry lowering is considered by displacing $H$ atoms away 
from positions given by the $HoD_{3}$ structure of $YH_{3}$.  \cite{Kelly}

Other models\cite{Ng,Eder}, based on strong electron correlations, 
have been proposed to explain the metal insulator transition.  Ng et 
al.  \cite{Ng} studied the effect of correlations on the bandwidth of 
hydrogen induced states.  Hydrogen is present in the form of $H^{-}$ 
as determined from electromigration experiments \cite{elmigr}, where 
one electron is taken from the metal host.  The two electrons on 
$H^{-}$ are strongly correlated and the essence of the result of Ng et 
al.  \cite{Ng} is that the opening of the band gap is due to a 
correlation-induced band narrowing.

The model of Eder et al. 
\cite{Eder} is based on the observation that the radius of the 
hydrogen is very sensitive to the occupation number. 
The two electrons on $H^{-}$ are correlated but with drastically 
different radii around the proton. This results in a 
so-called breathing mode and a local singlet-like bound state with one 
electron on the proton and the other on the neighboring metal orbitals. 
Already at the meanfield 
level this introduces a significant correction to the potential at the 
$H$ site, effectively increasing the electron affinity, 
lowering the hydrogen band and opening the gap. 
There is an interesting and appealing connection 
between this model and the Zhang-Rice singlet \cite{ZRsin} in  
high temperature superconductors (HTc's). 
The two electrons form a singlet analogous to holes in the HTc's.  

On the 
other hand, very recent so-called GW-calculations \cite{Miyake,vanG} demonstrate 
the formation 
of a sufficient gap to explain the metal-insulator transition without 
need of strong electron correlations. Rather, these calculations indicate 
that the gap opening is  described as in normal 
semiconductors. For semiconductors LDA does not produce the 
correct gap whereas the 
self-energy corrections included in the GW-calculations are able to 
account for this deficiency. 

Indeed, detailed angle-resolved photoemission (ARPES) 
experiments are needed to favor one or the other model.  
However, 
practically all previous work on metal hydrides has been done on 
polycrystals and/or on samples that are capped with a protective Pd 
layer.  In order to perform ARPES experiments, uncapped single 
crystalline material is needed.  Furthermore, preparation has to take 
place \emph{in situ} since $Y$ is extremely 
reactive.  

Here we present, to our knowledge, 
the first ARPES data on the trihydride phase of uncapped single 
crystalline films.
We find that the overall bandwidth of the Y trihydride phase agrees 
with LDA calculations. However, a set of flat bands is observed with 
significantly higher binding energy, a fact that argues in favor of 
ideas of the model proposed by Eder et al. \cite{Eder}.

%
%

Experiments were performed in a Vacuum Generators ESCALAB Mk II 
spectrometer with motorized sequential angle-scanning data 
acquisition. \cite{osterw91e}  ARPES measurements were 
performed with monochromatized He I$\alpha$ ($h\nu=$21.2 eV).  The setup 
including a plasma discharge lamp with monochromator is 
described elsewhere \cite{Pillo98e}.  The energy resolution is 35 meV 
for He I$\alpha$ measurements.  All measurements have been done with 
the sample kept at room temperature.  The substrate used to grow on
the single crystalline $Y$ hydride films was a W(110) single crystal. 
Details of the set-up used, the preparation and the careful 
characterization of the 
crystal and electronic structure and the calibration of the 
$H$-concentration has been described elsewhere \cite{hayozAuto,hayoz98}. 
In brief, this 
set-up allows to prepare clean, single-crystalline rare-earth hydride 
films using hydrogen pressures up to 1.3 bar.  It combines a 
high-pressure reaction cell
with a custom made hydrogen 
purification system based on a Pd-24\%Ag permeation tube and a sorption pump, 
i.e., a getter alloy (70\%Zr-24.6\%V-5.4\%Fe).  The hydrogen 
purification system removes residual contaminations from 6N $H_{2}$ 
gas efficiently.  The overall design is such that the sample never 
gets in contact with non ultra-high-vacuum compartments. 
The results presented here are taken from 200 {\AA} thick, well-ordered single 
crystalline films\cite{hayozAuto} in the trihydride phase.  The hydrogen 
composition is determined via photoelectron diffraction and X-ray 
photoelectron spectroscopy (not shown) to be $YH_{2.9}$. \cite{hayozAuto}
\begin{figure}[t!]
	\centerline{\epsfig{file=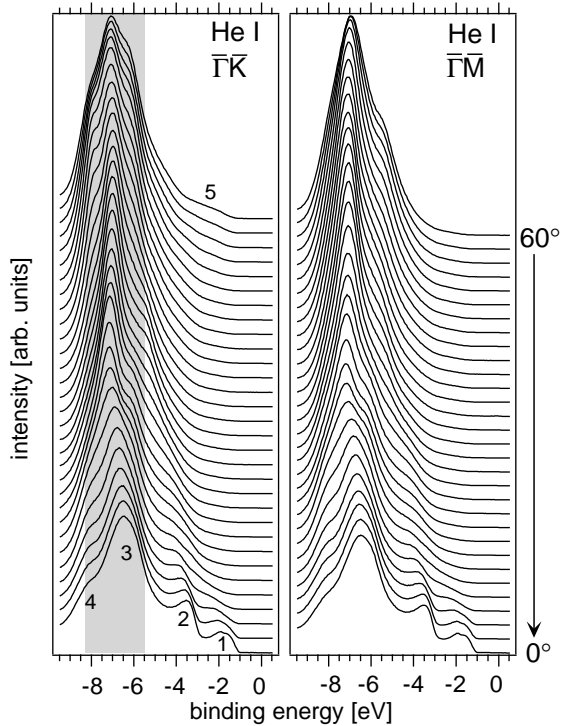,width=7.6cm}}
	\vspace{0.3cm} 
	\caption{
	Energy distribution curves taken along the 
	$\bar{\Gamma} \bar{K}$ (left) 
	and $\bar{\Gamma} \bar{M}$ (right) directions of the 
	surface Brillouin zone (see Fig. 2) using HeI 
	($h\nu$ = 21.2 eV) radiation and collecting spectra up to $60\rm^{o}$ 
	off-normal emission. Normal emission ($0\rm^{o}$) corresponds to 
	$\bar{\Gamma}$; the $\bar{K}$ and $\bar{M}$ points are reached 
	approximately at 
	$17\rm^{o}$ and $15\rm^{o}$ off-normal emission, respectively, for 
	$0$ eV binding energy. Mapping into k-space is shown in Fig. 4. A 
	shaded area marks weakly dispersing, broad spectral features (see 
	text).
	}
	\label{fsm}
\end{figure}

Bandstructure calculations using the full-potential 
linearized augmented plane-wave method \cite{blaha90e} within the 
generalized-gradient approximation \cite{perdew96e} 
have been performed for comparison with 
the experiments. 
Calculations were performed for the 
$HoD_{3}$ structure with space group 165 ($P\bar{3}c1$). The lattice 
parameters used for the $YH_{3}$ calculation 
were $a=b=6.34 \ $\AA    \ and   $c=6.6 \ $\AA. A total 
of 485 k-points within the irreducible wedge of the Brillouin zone (BZ) were 
considered for the self consistency cycles and convergence was reached 
to within 0.1 mRy.

%
%
Figure 1 shows ARPES data measured along two high symmetry directions 
of the surface BZ. The situation in 
reciprocal space is illustrated in Fig.  2 where the surface BZ (left) 
and a cut through the extended bulk Brillouin zone containing the 
$\Gamma, A, H, L, M, K$ high symmetry points (right) is shown.  
$YH_{3}$ with the $HoD_{3}$ structure consists of a $(\sqrt{3} \times 
\sqrt{3})-R30\rm^{o}$ reconstruction with respect to the hexagonal 
$Y$-lattice, 
i.e., the unit cell has $a$ and $b$ vectors which are $\sqrt{3}$ times 
longer and rotated by $30\rm^{o}$.  Therefore the $YH_{3}$ BZ is 
smaller and rotated accordingly (Fig.  2).

Inspecting the experimental spectra (Fig.1 ) a similar behavior is 
observed for both directions.  The dispersion appears relatively weak 
and the electronic states do not reach the Fermi level ($E_{F}$) or 
$0$eV binding energy, which is indicative for a gap. 
The $d$-states that are present in $Y$ ([Kr]$4d^{1}5s^{2}$) and 
$YH_{2}$ (not shown) have disappeared.\cite{JHDiss}  
Five dispersing features are easily discernible and are labeled from 
$1$ to $5$.  A broad almost dispersionless maximum between 5.5 and 8.3 
eV (shaded area) persists for all angles respectively $k$-points.  At first sight one is 
tempted to interpret the fact that the dispersion is weak and the 
spectral features are broad as caused by strong electron correlations and 
self-energy effects.
\begin{figure}[]
	\centerline{\epsfig{file=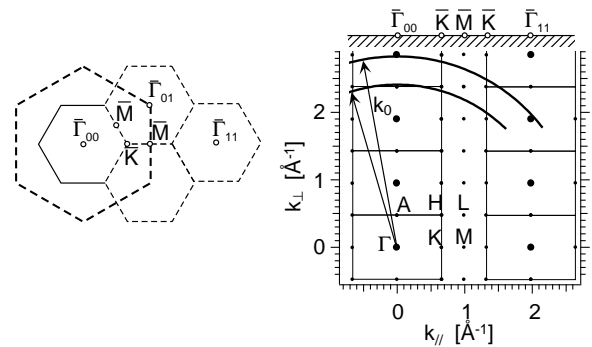,width=8.6cm}}
	\vspace{0.3cm}
	\caption{
	Situation in k-space. (Left) The surface Brillouin zone for the $HoD_{3}$
	structure with high symmetry points. The large hexagon represents the 
	surface Brillouin zone for the Y sublattice. (Right) Section across 
	the bulk Brillouin zone containing the $\Gamma, A, H, L, M, K$ high 
	symmetry points. Spherical segments indicate the k-space 
	region probed within the free-electron final state approximation used 
	for the model calculation in Fig. 4.   
	}
	\label{leed}
\end{figure}

However, taking a closer look reveals a different point of view.  In 
Fig.  3 the result of our bandstructure calculation of $YH_{3}$ in the 
$HoD_{3}$ structure is displayed.  As mentioned above the approach 
using LDA does not show a gap and the system is metallic.  The 
calculation presented here is in good agreement with the published 
results of Kelly et al.\cite{Kelly}.  We notice that there are many 
bands.  This is a consequence of the large real-space unit cell of the 
$HoD_{3}$ structure containing many atoms.  In fact, there are 24 
atoms including 6 yttrium atoms and 18 hydrogen atoms in the unit 
cell.  The important point to notice here is that there is a set of 
rather flat, hydrogen-induced bands (as deduced from a band character 
analysis) between 1.8 and 4.6 eV (dark shaded area) extending over the 
whole bulk BZ. 
Such a set of flat bands naturally gives rise to a high density 
of states in this energy region and leads to the interpretation of the broad, 
high intensity features in the experiment (shaded area in 
Fig.  1) in terms of these flat bands.
The energy width of the region 
in the experiment agrees well with the energy interval of the flat 
bands in the bandstructure calculation.  The only difference is that 
the experimental bands occur at a 3.7 eV greater binding energy.
\begin{figure}[t]
	\centerline{\epsfig{file=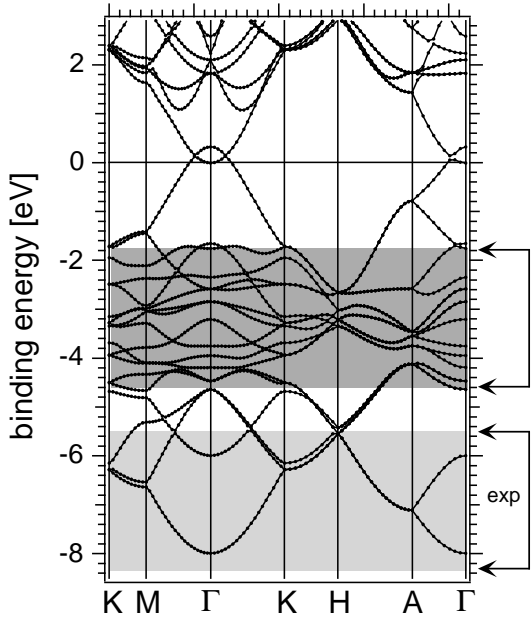,width=7.0cm}}
	\vspace{0.3cm} 
	\caption{
	LDA based bandstructure calculation for the $HoD_{3}$ structure. 
	Indicated is the region (dark shading) of flat, hydrogen-induced
	 bands and the 
	corresponding region (light shading) where the experiment (Fig. 1) 
	shows a high density of states. 
	(see text)
	 }
	\label{vb}
\end{figure}

In order to obtain a more detailed comparison we use a gray scale 
representation of both, the experimental data and the calculated band 
structure.  Figure 4 displays the results along the 
$\bar{\Gamma}   
\bar{K}$-direction.  For the experimental data (left) the second derivative 
of the spectra of Fig.  1 have been plotted as a function of the wave 
vector parallel to the surface.  The reason for plotting the second 
derivative is to flatten the spectra and to accentuate the dispersing 
features. The calculation (right) follows the 
free electron final state wave vectors drawn in Fig.  2 \cite{calc}.  
Again we can 
identify the flat bands (marked by shading the energy scales) 
in both the experiment and the calculation 
where they occur with a 3.7 eV smaller binding energy.  In addition we 
can identify the bands labeled $1$ to $5$ as already indicated in Fig.  
1.  Therefore comparing our experiments with the LDA calculation shows 
that the overall band width compares well.  The differences are a rigid 
shift of approximately 1.5 eV and the position of the flat bands.  
However, a detailed comparison of different bands and their relative 
positions depends on the validity of the free-electron final-state 
approximation.  Therefore we cannot, at present, give a robust 
statement about the exact position of the top of valence band (label 
1, Fig.  4, left) and, as a consequence, about the size of the gap.  
This depends on the exact location in the component of the wave vector 
perpendicular to the surface, $k_{\perp}$.  On 
the other hand, the difference in the position of the flat bands is 
independent of this approximation since the flat bands extend over the 
whole BZ as seen from experiment (Fig.  1) and theory (Fig.  3).
\begin{figure}[]
	\centerline{\epsfig{file=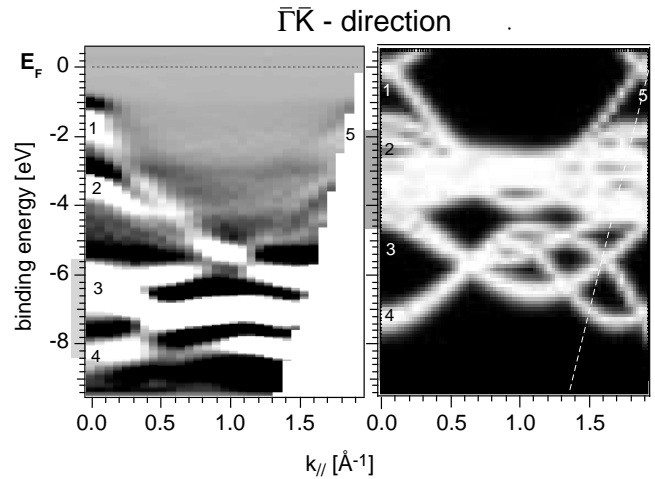,width=8.6cm}}
	\vspace{0.3cm}
	\caption{
	Second derivative of spectra from Fig. 1 mapped into k-space using 
	gray scale intensity coding with high intensity in white (left). Model 
	calculation
	 using 
	the 
	free electron final state approximation (right) (see text). 
	}
\end{figure}

It is interesting now to compare our results with theoretical 
models. At first we look at the GW calculations. In the study of 
Miyake et al. \cite{Miyake} the gap opens due to a shifting up of the 
conduction bands whereas in the case of van Gelderen et al. 
\cite{vanG} the gap opens because of both, shifting up of the 
conduction 
bands and shifting down of valence bands. Otherwise the quasiparticle 
wavefunctions are practically identical to the LDA wave functions and 
there is basically a one to one correspondence between the bands of 
the LDA and GW bandstructures. The total band widths do not differ 
significantly as well. From our experiments we conclude that the 
overall bandwidth is well reproduced by LDA and GW-calculations. 
However, the high binding energy of the flat bands is not reproduced. 
Even the downwards 
shift of the valence bands ($\approx $1 eV) in the case of van Gelderen 
et al. \cite{vanG} 
is not sufficient and does not account for the $relative$ shift of 
certain bands. Nevertheless, from the comparison in Fig. 4 (therefore assuming 
the free-electron final-state to be valid) we notice that a downward 
shift of the LDA results, as would be obtained for the GW-calculation 
produces quite a good agreement if we do not consider the flat bands. 

Clearly, the good agreement of the overall bandwidth with the LDA 
calculations is in contrast to the model of Ng et al. \cite{Ng} where 
strong electron correlations induce a band narrowing. 
However, we cannot exclude that individual bands are subject to 
significant narrowing. The reason is that from the experimental data, 
basically due to the large 
number of bands, it is not possible to identify every individual band. 
The width of the experimentally observed spectral features is not 
resolution limited and might be explained either by self-energy 
effects or also by a relatively high mobility of hydrogen atoms and 
defects due to the substoichiometry of the films. Furthermore, there is 
still debate on whether the $HoD_{3}$ structure really corresponds to 
the one for $YH_{3}$. \cite{udo,vangel} 

Finally we compare our results to the model by Eder et al.\cite{Eder}.  
The dependence of the occupation number of electrons on the hydrogen 
site is incorporated in their model Hamiltonian by having a novel 
hopping integral.  The resulting singlet-like bound state is manifest 
via increasing the binding energy of the hydrogen states.  This model 
and the one by Ng et al.  \cite{Ng} have in common that they 
treat hydrogen as impurities in the background of the metal atoms.  
However, whereas in the treatment of Ng et al.  the result is a 
narrowing of bands, Eder et al.\cite{Eder} predict a shift of the 
potential at the hydrogen site $retaining$ the broad hydrogen band.

It is now very tempting to interpret our 
data, showing a shift of bands towards higher binding energy, in terms 
of such an additional potential on hydrogen sites. Observing Fig. 3 
it appears that only the set of flat bands is strongly shifted down 
whereas the others, covering the full bandwidth, are consistent 
with the GW approach. This might be interpreted in the sense that 
the bands within the set of flat bands have a more localized 
character and therefore are stronger affected by the singlet-like 
bound state. However, a detailed comparison of how different bands are 
affected is not possible since a realistic bandstructure for the 
$HoD_{3}$ structure is not available within Eder's model Hamiltonian.   

%
In conclusion we have performed ARPES experiments on single 
crystalline films of Y in the trihydride phase. A comparison with LDA 
calculations shows a good agreement of 
the overall bandwidth. No significant band narrowing is observed 
although a narrowing of individual bands cannot be excluded. However a 
set of hydrogen induced, flat bands are identified in the 
experiment with a binding energy $\approx$ 3.7 eV higher than in the 
calculation. Furthermore, based on the free-electron final-state 
approximation, we infer a rigid shift of the valence band towards 
higher binding energy by more than 1 eV which is consistent with 
GW-calculations \cite{vanG} and the opening of a gap. The downwards shift of flat 
bands supports the model proposed by Eder et al.\cite{Eder} 
predicting a shift of the potential at the hydrogen site and $retaining$ 
the broad hydrogen band.

Skillful technical assistance was provided 
by E. Mooser, O. Raetzo, R. Schmid, O. Zosso, Ch.  Neururer and F. 
Bourqui.  This project has been supported by the Fonds National Suisse 
de la Recherche Scientifique.  


\end{multicols}

\begin{thebibliography}{10}
\bibitem{calc} The gray scale levels of the calculation are 
determined by checking energy conservation with respect to the free-
electron final-state, i.e., perfect energy conservation appears in 
white. An inner potential of 10 eV and a workfunction of 3 eV have 
been assumed. No matrix-elements for the photoemission process are considered.
\end{thebibliography}
\end{document}